\documentclass[final,5p,times,twocolumn]{elsarticle}

\usepackage{amssymb}
\usepackage{amsmath}
\usepackage{url}
\journal{Expert Systems with Applications}

\begin{document}
\begin{frontmatter}
	\title{ParamNet: A Dynamic Parameter Network for Fast Multi-to-One Stain Normalization}
%		\author[smu]{Hongtao \snm{Kang} }
%		\author[hust]{Die \snm{Luo} }
%		\author[tongji]{Li \snm{Chen} }
%		\author[sfy]{Junbo \snm{Hu} }
%		\author[hust]{Tingwei \snm{Quan} }
%		\author[hust]{Shaoqun \snm{Zeng} }
%		\author[smu]{Shenghua \snm{Cheng} \corref{cor2}}
%		\ead{chengsh2023@smu.edu.cn}
%		\author[hust]{Xiuli \snm{Liu} \corref{cor1}}
%		\ead{xlliu@mail.hust.edu.cn}
\author[smu]{Hongtao Kang }
\author[hust]{Die Luo }
\author[tongji]{Li Chen }
\author[sfy]{Junbo Hu }
\author[hust]{Tingwei Quan }
\author[hust]{Shaoqun Zeng }
\author[smu]{Shenghua Cheng \corref{cor2}}
\ead{chengsh2023@smu.edu.cn}
\author[hust]{Xiuli Liu \corref{cor1}}
\ead{xlliu@mail.hust.edu.cn}
	
	\address[smu]{School of Biomedical Engineering and Guangdong Provincial Key Laboratory of Medical Image Processing, Southern Medical University, Guangzhou, Guangdong 510515, China}
	
	\address[hust]{Britton Chance Center and MoE Key Laboratory for Biomedical Photonics, Wuhan National Laboratory for Optoelectronics-Huazhong University of Science and Technology, Wuhan, Hubei 430074, China}
	
	\address[tongji]{Department of Clinical Laboratory, Tongji Hospital, Tongji Medical College, Huazhong University of Science and Technology, Wuhan, Hubei 430030, China}
	
	\address[sfy]{Department of Pathology, Hubei Maternal and Child Health Hospital, Wuhan, Hubei 430072, China}
	
	\cortext[cor1]{Corresponding author at: Britton Chance Center and MoE Key Laboratory for Biomedical Photonics, Wuhan National Laboratory for Optoelectronics-Huazhong University of Science and Technology, Wuhan, Hubei 430074, China}
	\cortext[cor2]{Corresponding author at: School of Biomedical Engineering and Guangdong Provincial Key Laboratory of Medical Image Processing, Southern Medical University, Guangzhou, Guangdong 510515, China}
	%\author{Hongtao Kang, Die Luo, Li Chen, Junbo Hu, Shenghua Cheng, Tingwei Quan, Shaoqun Zeng and Xiuli Liu
		%\thanks{This work is supported by the NSFC project (grant 62375100, 62201221), China Postdoctoral Science Foundation (grant 2021M701320, 2022T150237) and the Director Fund of Wuhan National Laboratory for Optoelectronics. \textit{(Corresponding author: Xiuli Liu and Li Chen.)}}
		%\thanks{H Kang, D Luo, T Quan, S Zeng and X Liu are with the Britton Chance Center for Biomedical Photonics, Wuhan National Laboratory for Optoelectronics, Huazhong University of Science and Technology, Wuhan, Hubei 430074, China, and also with the MoE Key Laboratory for Biomedical Photonics, School of Engineering Sciences, Huazhong University of Science and Technology, Wuhan, Hubei 430074, China(e-mail: khtao@hust.edu.cn; luodie@hust.edu.cn; quantingwei@hust.edu.cn; sqzeng@mail.hust.edu.cn; xlliu@mail.hust.edu.cn).}
		%\thanks{S Cheng is with School of Biomedical Enginering and Guanodong Provincial Key Laboratory of Medical Image Processing, Southern Medical University, Guanqzhou, Guangdong 510515, China (e-mail: 905806158@qq.com).}
		%\thanks{L Chen is with Department of Clinical Laboratory, Tongji Hospital, Huazhong University of Science and Technology, Wuhan, Hubei 430030,China(email:chenliisme@126.com).}
		%\thanks{J Hu is with Department of Pathology, Hubei Maternal and Child Health Hospital, Wuhan, Hubei 430072, China(email:cqjbhu@163.com).}
		%}
	
	\begin{abstract}
		
		%In practice, digital pathology images are often affected by various factors, resulting in very large differences in color and brightness. Stain normalization can effectively reduce the differences in color and brightness of digital pathology images, thus improving the performance of computer-aided diagnostic systems. Conventional stain normalization methods rely on one or several reference images, but one or several images may not adequately represent the entire dataset. Although learning-based stain normalization methods are a general approach, they use complex deep networks, which not only greatly reduce computational efficiency, but also risk introducing artifacts. StainNet uses three layers of 1×1 convolutions to perform normalization, which is too simple to perform many-to-one staining normalization. In this study, we introduced dynamic-parameter network based on StainNet and proposed a novel method for stain normalization, called ParamNet. ParamNet addresses the challenges of limited network capacity and computational efficiency by introducing dynamic parameters (weights and biases of convolutional layers) into the network design. By effectively leveraging these parameters, ParamNet achieves superior performance in stain normalization while maintaining computational efficiency. Results show ParamNet can normalize one whole slide image (WSI) of 100,000×100,000 within 25s. The code is available at: https://github.com/khtao/ParamNet.
		
		In practice, digital pathology images are often affected by various factors, resulting in very large differences in color and brightness. Stain normalization can effectively reduce the differences in color and brightness of digital pathology images, thus improving the performance of computer-aided diagnostic systems. Conventional stain normalization methods rely on one or several reference images, but one or several images may not adequately represent the entire dataset. Although learning-based stain normalization methods are a general approach, they use complex deep networks, which not only greatly reduce computational efficiency, but also risk introducing artifacts.  Some studies use specialized network structures to enhance computational efficiency and reliability, but these methods are difficult to apply to multi-to-one stain normalization due to insufficient network capacity.  In this study, we introduced dynamic-parameter network and proposed a novel method for stain normalization, called ParamNet. ParamNet addresses the challenges of limited network capacity and computational efficiency by introducing dynamic parameters (weights and biases of convolutional layers) into the network design. By effectively leveraging these parameters, ParamNet achieves superior performance in stain normalization while maintaining computational efficiency. Results show ParamNet can normalize one whole slide image (WSI) of 100,000×100,000 within 25s. The code is available at: \url{ https://github.com/khtao/ParamNet}.
	\end{abstract}
	
	\begin{keyword}
		Stain normalization, Cytopathology, Histopathology, Convolutional neural network, Generative adversarial network.
	\end{keyword}
\end{frontmatter}
\section{Introduction}
\label{sec:introduction}
With the development of computer technology, computer-aided diagnosis (CAD)  has received more and more attention, and many automated pathology diagnosis tools have been developed \citep{Advances,zhang2014survey,cheng2021robust,jo2021comparative}. In practice, however, the pathology images from different medical centers are often very different due to the differences in staining reagent manufacturers, staff operations, and imaging systems \citep{vahadane2016structure,adversarial}. Even within one medical center, there is some variability over time. For pathologists, small color differences will not affect the diagnosis, but larger color differences will affect even pathologists \citep{Ismail707}. For CAD systems, even small color differences can affect their reliability \citep{Ciompi}. Stain normalization can effectively reduce these differences by normalizing different color distributions to a uniform color distribution, thereby improving the performance of the CAD system \citep{Cai,anghel2019high,khan2014nonlinear}.

Conventional stain normalization methods rely on one or several reference images to extract color mapping relationships. However, one or several images may not adequately represent the entire dataset \citep{zheng2020stain,kang2021stainnet}. Conventional methods can be mainly divided into two categories, stain matching based methods \citep{reinhard2001color} and stain-separation methods \citep{ruifrok2001quantification,macenko2009method,vahadane2016structure,salvi2020stain,gupta2020gcti}. Stain matching based methods usually match the mean and standard deviation of the source image with the target image in the color space, e.g. LAB \citep{reinhard2001color}. Stain-separation methods try to separate and normalize each staining channel independently in Optical Density (OD) space \citep{ruifrok2001quantification,macenko2009method,vahadane2016structure,salvi2020stain,gupta2020gcti}. It is very difficult to separate each staining channel for some pathology images using multiple stains, such as cytopathology images \citep{kang2021stainnet,gill2012cytopreparation}.

With the rise of deep learning, more and more stain normalization methods have adopted deep learning-based methods \citep{cho2017neural,chen2021unsupervised,salehi2020pix2pix,tellez2019quantifying,zanjani2018stain,shaban2019staingan,zhou2019enhanced,zhu2017unpaired,shrivastava2021self}. These methods can be roughly divided into two categories, one is the pix2pix-based methods \citep{cho2017neural,chen2021unsupervised,salehi2020pix2pix,tellez2019quantifying,zanjani2018stain}, another is CycleGAN-based methods \citep{shaban2019staingan,zhou2019enhanced,zhu2017unpaired,shrivastava2021self}.

The pix2pix-based methods try to reconstruct the original image from the color transformed image. These color transformation methods include grayscale \citep{cho2017neural,chen2021unsupervised,salehi2020pix2pix}, Hue-Saturation-Value (HSV) \citep{tellez2019quantifying} and CIEL*a*b* color space \citep{zanjani2018stain}. However, the color variation of the source image still affects the reconstructed image. The CycleGAN-based methods generally adopt cycle consistency loss and adversarial loss to learn the color mapping from the real source and target image \citep{shaban2019staingan,zhou2019enhanced,zhu2017unpaired,shrivastava2021self}. By using real source and target images, the image normalized by these methods is very close to the real target image. However, in real data, due to the mismatch in the distribution of the source and target images and the instability of the GAN network, artifacts, abnormal brightness, or distortion may occur \citep{zheng2020stain,kang2021stainnet,lei2020staincnns}. These phenomena are unacceptable in stain normalization. Some researchers have tried to solve this problem by using additional structures to make the source image and the normalized image consistent in content \citep{mahapatra2020structure,lahiani2019perceptual}. These methods include using the segmentation branch to ensure semantic consistency \citep{mahapatra2020structure} and using additional branches to ensure the content consistency\citep{lahiani2019perceptual}. However, these methods add extra branches, complicate training, and cannot really guarantee against anomalies.

Because pathology images often contain a large number of pixels, researchers hope to improve the computational efficiency of stain normalization algorithms. Anand et al. \citep{anand2019fast} optimized the Vahadane \citep{vahadane2016structure}. and Anghel et al. \citep{anghel2019high} optimized the Macenko \citep{macenko2009method} method. Zheng et al. \citep{zheng2020stain} proposed to use 1×1 convolution and sparse routing to achieve fast stain normalization. However, the style of the normalized image by this method is formed automatically and cannot be normalized to a specific color distribution. StainNet \citep{kang2021stainnet} used distillation learning and 1×1 convolutional neural network to reduce the amount of normalization computation. Benefiting from its fully 1×1 convolutional network structure, StainNet  \citep{kang2021stainnet}  can retain as much information from the source image as possible. However, its overly simplified network structure limits its color mapping capabilities. When images of multiple styles need to be normalized into one style,  StainNet \citep{kang2021stainnet} must be trained multiple times to obtain different mapping relationships, which undoubtedly increases its application cost.

In this study, we present ParamNet, a novel method to address the challenges of limited network capacity and computational efficiency by introducing dynamic parameters into the network design. These parameters are carefully designed to enhance the network's representation capabilities, enabling it to effectively handle stain variations in histopathological and cytopathological images. ParamNet achieves superior performance in stain normalization while maintaining computational efficiency, making it a promising solution for practical applications in histopathology and cytopathology. 

Our ParamNet contains two sub-networks, a modified ResNet18 \citep{he2016deep} as the prediction sub-network and a fully 1×1 convolutional network as the color mapping sub-network, where the prediction sub-network can predict all parameters (weights and biases of convolutional layers) of the color mapping sub-network at low resolution, and the color mapping sub-network can directly perform stain normalization at the original resolution. The fully 1×1 convolutional color mapping sub-network contains only two convolutional layers with a total of 59 parameters (weights and biases of convolutional layers), which allows our network to be computationally efficient and structurally consistent. The prediction sub-network running at low resolution can automatically determine the parameters (weights and biases of convolutional layers) of the color mapping sub-network according to each input image. So, our ParamNet can perform various stain normalization tasks, including one-to-one stain normalization and multi-to-one stain normalization, in a fast and robust manner. This allows our ParamNet to normalize an 100,000×100,000 whole slide image (WSI) within 25s. Our adversarial training framework introduces a texture module between ParamNet and the discriminator, which solves the misfit between weak generators and strong discriminators. Further, we validate the effectiveness of the method on datasets containing multiple staining styles.
\begin{figure*}[!t]
	\footnotesize 
	\centering
	\includegraphics[width=154 mm]{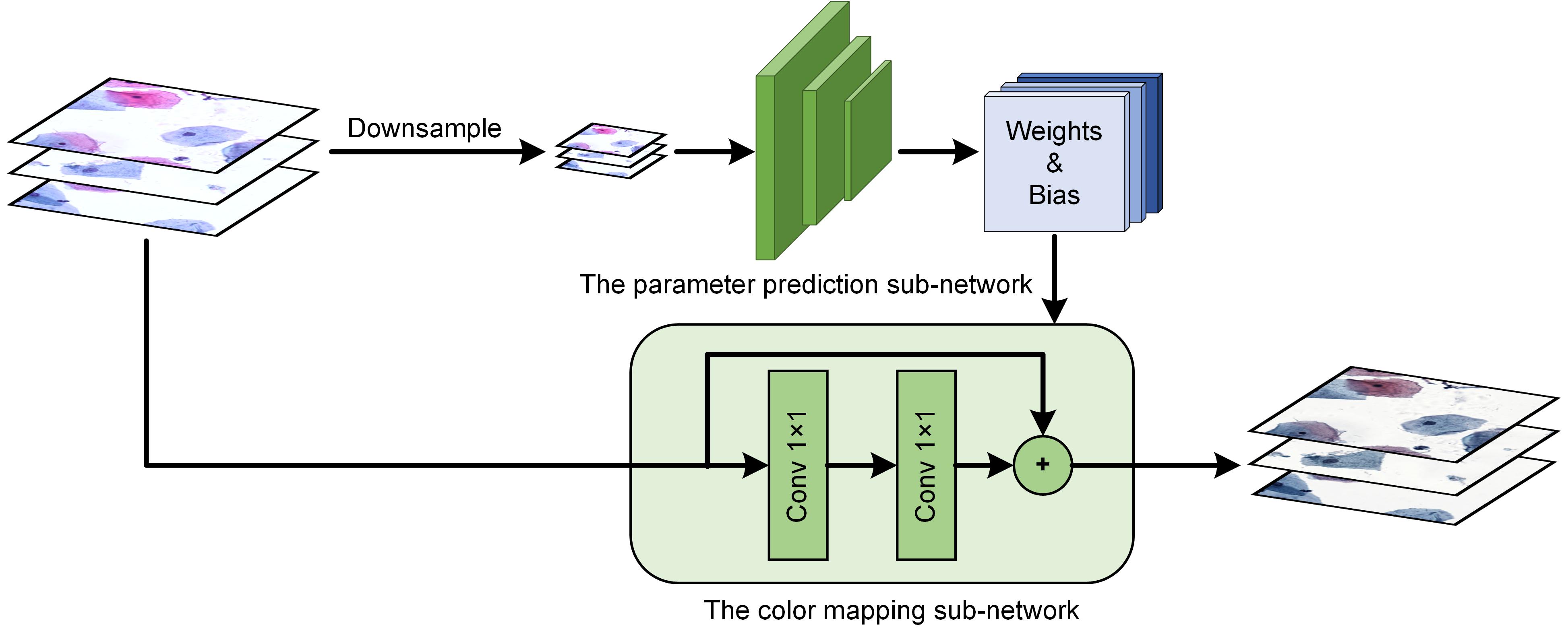}
	\caption{Network structure of ParamNet. First, the parameter prediction sub-network predicts the parameters (weights and biases of convolutional layers) of the color mapping sub-network at low resolution. Then, the color mapping sub-network uses these parameters to normalize the input image at original resolution.}
	\label{fig1}
\end{figure*}

\section{Methods}
In this section, we introduce the network structure of ParamNet and its adversarial training framework.

\subsection{Network Structure of ParamNet}

Stain normalization is an important preprocessing step in CAD system. A practical stain normalization method should have the following features: 1. Fast: Pathological full-slide images usually contain a large number of pixels, so the stain normalization method needs to be computationally efficient. 2. Structural consistency: The stain normalization method should have the same structure as the source image and only change the color style. 3. Multi-domain: The stain normalization method should be able to normalize the images of multiple color styles at the same time.

StainNet \citep{kang2021stainnet} is a fast and robust stain normalization method, but only for one-to-one stain normalization. When normalizing multiple color styles, StainNet \citep{kang2021stainnet} has to be trained multiple times to obtain different network parameters (weights and biases of convolutional layers). To solve this problem, we propose ParamNet, which can automatically determine the appropriate network parameters (weights and biases of convolutional layers) for each input image.

As shown in Fig. \ref{fig1}, ParamNet consists of two parts, one is the parameter prediction sub-network, and another is the color mapping sub-network. The parameter prediction sub-network can predict all the parameters (weights and biases of convolutional layers) of the color map sub-network according to the color style of the input image at low resolution. The color mapping sub-network utilizes these parameters to directly normalize the source image at original resolution. The feature of dynamic parameter ensures that our network has a sufficient capability for multi-domain normalization.

In our implementation, the parameter prediction sub-network uses a modified ResNet18 \citep{he2016deep}, in which all BatchNorm layers are removed and all convolutional layers are set as biased convolutional layers. For the stability of the network, the output of the prediction sub-network will be normalized by the tanh function, and then multiplied by a learnable parameter. The learnable parameter represents the numerical range of the parameters (weights and biases of convolutional layers) in the color mapping sub-network and its initial value is set to 4.5. The input image of the parameter prediction sub-network are downsampled to 128×128 by default. 

The color mapping sub-network is a fully 1×1 convolutional network, which is similar to StainNet \citep{kang2021stainnet}. Since only 1×1 convolutional layer is used, our color mapping sub-network only maps a single pixel without interference from the input image's local neighborhood, thus ensuring the structural consistency of the output image. The rectified linear unit (ReLU) is used as the activation function of the convolutional layer, and tanh is used as the activation function of the output layer. Due to the dynamic network, the color mapping sub-network can use a more concise network structure. The color mapping sub-network uses two 1×1 convolutional layers by default, and the number of channels is set to 8. So, the color mapping sub-network only contains 59 parameters, which allows our network to be computationally efficient.

\begin{figure*}[!t]
	\footnotesize 
	\centering
	\includegraphics[width=120 mm]{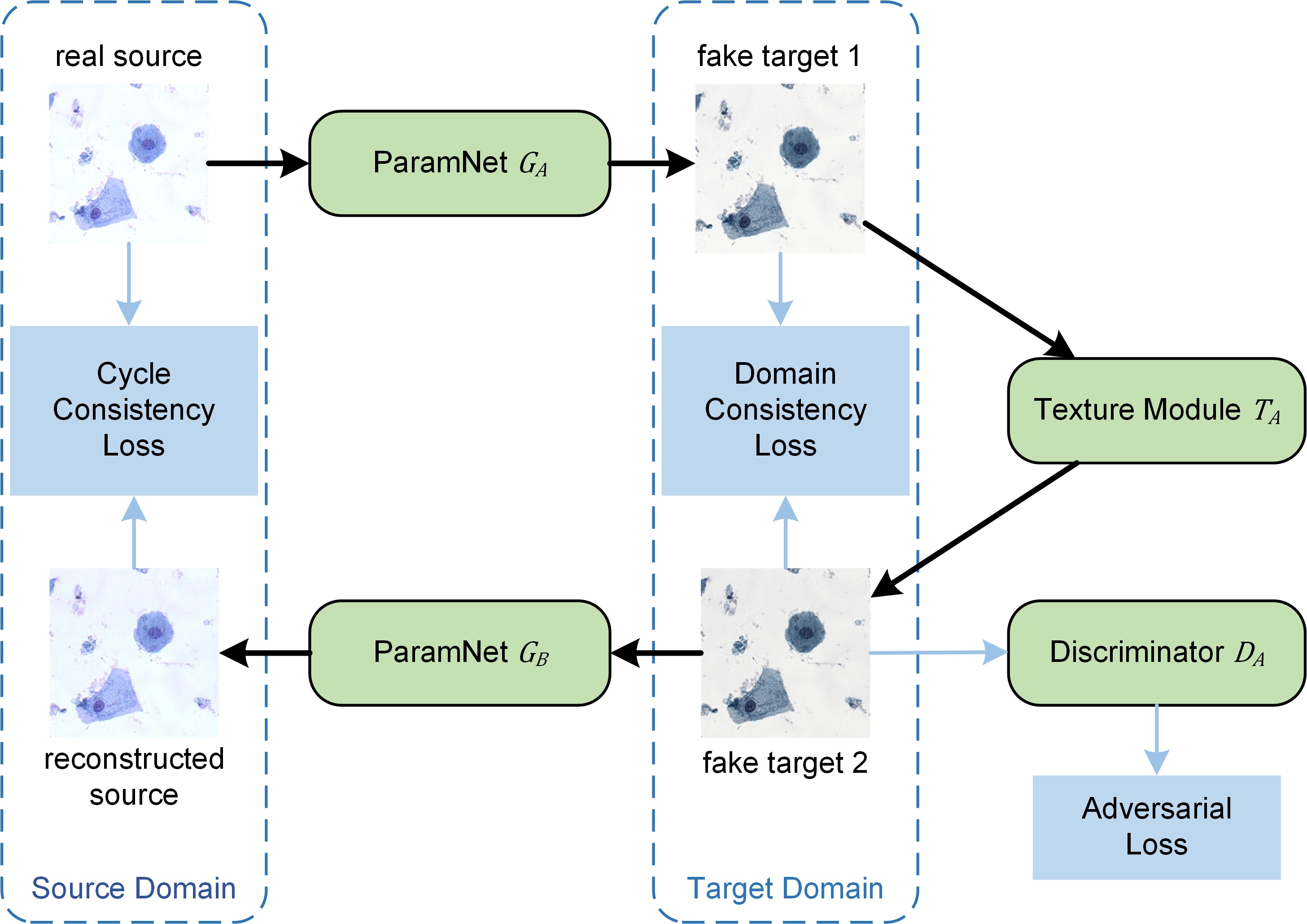}
	\caption{Adversarial training framework of ParamNet. Images from source domain are mapped to the target domain, after which texture details are added, and the result is mapped back to the source domain. The same reverse process is also performed for the images from the target domain. The difference is that in the inverse process, the texture module uses $T_B$ instead of $T_A$, and the discriminator uses $D_B$ instead of $D_A$.}
	\label{fig2}
\end{figure*}

\subsection{Adversarial Training Framework of ParamNet}
Since ParamNet only transforms the color value of the source image, it retains a lot of information of the source image. In adversarial training, the discriminator can easily distinguish the difference between the image normalized by ParamNet and the real image, resulting in the failure of generating adversarial training. To solve this problem, we introduce a texture module to transform the texture of the output image of ParamNet to obtain an image that can deceive the discriminator. Further, during training, we randomly transform image resolutions (0.5$\sim$1.0) to stabilize learning.

As shown in Fig. \ref{fig2}, there are two generators (ParamNet A ($G_A$) and ParamNet B ($G_B$)), two texture modules (Texture Module A ($T_A$) and Texture Module B ($T_B$)) and two discriminators (Discriminator A ($D_A$) and Discriminator B ($D_B$)) included in our framework. $G_A$ is used to map the images from source domain to target domain, and $G_B$ is used to map the images from target domain to source domain. $T_A$ is used to add some details to the normalized image by $G_A$. $T_B$ is used to add some details to the normalized image by $G_B$. $D_A$ is used to distinguish the difference between the real target image and the output of $T_A$, and $D_B$ is used to distinguish the difference between the real source image and the output of $T_B$.

There are four losses included in our framework, adversarial loss, cycle consistency loss, domain consistency loss and identity loss. 
%The total objective function can be expressed as: \begin{equation*} \begin{split} \mathcal { L } = \mathcal {L}_{GAN}(G_A,T_A,D_A) + \mathcal {L}_{GAN}(G_B,T_B,D_B) \\ + \mathcal {L}_{cyc}(G_A,G_B,T_A,T_B) + \mathcal {L}_{dom}(G_A,G_B,T_A,T_B) \\+ \mathcal{L}_{ide}(G_A,G_B,T_A,T_B), \end{split} \tag{1} \end{equation*}
%where $\mathcal {L}_{GAN}(G_A,T_A,D_A)$ and $\mathcal {L}_{GAN}(G_B,T_B,D_B)$ are the adversarial loss, $\mathcal {L}_{cyc}(G_A,G_B,T_A,T_B)$ is the cycle consistency loss, $\mathcal {L}_{dom}(G_A,G_B,T_A,T_B)$ is the domain consistency loss and $\mathcal{L}_{ide}(G_A,G_B,T_A,T_B)$ is the identity loss.

The adversarial loss is used to ensure that the outputs of the two texture modules have the same distribution as the real source and target images. The objective function is defined as:
\begin{equation*} \begin{split}
		\mathcal {L}_{GAN}(G_A,T_A,D_A) = \mathbb{E}_{ t\sim p_{data}(t)}[\Vert D_A(t) \Vert_2] \\ + \mathbb{E}_{ s\sim p_{data}(s)}[\Vert 1-D_A(T_A(G_A(s))) \Vert_2]
	\end{split}  \tag{2}
\end{equation*}
\begin{equation*} \begin{split}
		\mathcal {L}_{GAN}(G_B,T_B,D_B) = \mathbb{E}_{ s\sim p_{data}(s)}[\Vert D_B(s) \Vert_2] \\ + \mathbb{E}_{ t\sim p_{data}(t)}[\Vert 1-D_B(T_B(G_B(t))) \Vert_2]
	\end{split}  \tag{3}
\end{equation*}
where $\Vert \cdot \Vert_2$ is the $\ell_{2}$-norm,  $s$ is the real source image, and $t$ is the real target image.

The cycle consistency loss constrains the consistency of image content by reconstructing the original image from the transformed image. The objective function can be expressed as:
\begin{equation*} \begin{split}
		\mathcal {L}_{cyc}(G_A,G_B,T_A,T_B) = \\ \mathbb{E}_{ s\sim p_{data}(s)}[\Vert s-G_B(T_A(G_A(s))) \Vert_1] \\ + \mathbb{E}_{ t\sim p_{data}(t)}[\Vert t-G_A(T_B(G_B(t))) \Vert_1]
	\end{split}  \tag{4}
\end{equation*}
where $\Vert \cdot \Vert_1$ is the $\ell_{1}$-norm.

Domain consistency loss aims to make the texture module change the image content as little as possible, and only adds some details that can deceive the discriminator. The objective function can be expressed as:
\begin{equation*} \begin{split}
		\mathcal {L}_{dom}(G_A,G_B,T_A,T_B) = \\ \mathbb{E}_{ s\sim p_{data}(s)}[\Vert G_A(s)-T_A(G_A(s)) \Vert_1] \\ + \mathbb{E}_{ t\sim p_{data}(t)}[\Vert G_B(t)-T_B(G_B(t)) \Vert_1]
	\end{split}  \tag{5}
\end{equation*}
where $\Vert \cdot \Vert_1$ is the $\ell_{1}$-norm.

Identity loss is designed such that when the input image comes from the corresponding target domain, the generator can maintain the identity map without making any changes to the image. Specifically, in this paper, we hope that when the inputs of $G_A$ and $T_A$ are from the target domain, $G_A$ and $T_A$ do not make any changes to the input. Similarly, when the inputs of $G_B$ and $T_B$ are from the source domain, $G_B$ and $T_B$ do not make any changes to the input. The objective function can be expressed as: 
\begin{equation*} \begin{split}
		\mathcal {L}_{ide}(G_A,G_B,T_A,T_B) = \\
		\mathcal {L}_{ide}(G_A,G_B)+
		\mathcal {L}_{ide}(T_A,T_B)
	\end{split}  \tag{6}
\end{equation*}
\begin{equation*} \begin{split}
		\mathcal {L}_{ide}(G_A,G_B) = \mathbb{E}_{ s\sim p_{data}(s)}[\Vert s-G_B(s) \Vert_1] \\ + \mathbb{E}_{ t\sim p_{data}(t)}[\Vert t-G_A(t) \Vert_1]
	\end{split}  \tag{7}
\end{equation*}
\begin{equation*} \begin{split}
		\mathcal {L}_{ide}(T_A,T_B) = \mathbb{E}_{ s\sim p_{data}(s)}[\Vert s-T_B(s) \Vert_1] \\ + \mathbb{E}_{ t\sim p_{data}(t)}[\Vert t-T_A(t) \Vert_1]
	\end{split}  \tag{8}
\end{equation*}

In our implementation, the network structure of the texture module and discriminator uses the generator and discriminator in StainGAN \citep{shaban2019staingan} respectively.

\section{Experiments and Results}
\subsection{Datasets}
In this study, we used four different datasets to verify the performance of different methods, including two public datasets and two private datasets. This study was approved by the Ethics Committee of Tongji Medical College, Huazhong University of Science and Technology. The related descriptions are below:
\subsubsection{Aligned Cytopathology Dataset}
The aligned cytopathology dataset from StainNet \citep{kang2021stainnet} was used to verify the similarity between the normalized and target images. In this dataset, the same slides (Thinprep cytologic test slides from Hubei Maternal and Child Health Hospital) were scanned using two different scanners. A scanner called Scanner O was equipped with a 20x objective lens with a pixel size of 0.2930 um. The other called the Scanner T was equipped with a 40x objective lens with a pixel size of 0.1803 um. In this dataset, the images from Scanner O and Scanner T are carefully aligned. There are 2,257 image pairs in the training set and 966 image pairs in the test set as shown in Table \ref{table1}. In this study, the images from Scanner O were used as the source images and the images from Scanner T were used as the target images.
\subsubsection{Aligned Histopathology Dataset}
The aligned histopathology dataset from StainNet \citep{kang2021stainnet} was used to verify the similarity between the normalized and target images. This dataset is originally from the publicly available part of the MITOS-ATYPIA ICPR'14 challenge \citep{mitosis}. This dataset was then aligned, registered, sampled and made publicly available by Kang et al. \citep{kang2021stainnet}. In this dataset, the same slides were scanned with two scanners (Aperio Scanscope XT called Scanner A and Hamamatsu Nanozoomer 2.0-HT called Scanner H). Among them, the training set contains 19,200 image pairs, and the test set contains 7,936 image pairs as shown in Table \ref{table1}. In this study, the images from Scanner A were used as the source images and the images from Scanner H were used as the target images.
\begin{table}[!h]
	\footnotesize 
	\centering
	\caption{ The number of image pairs on the aligned cytopathology and histopathology dataset.}
	\label{table1}
	%\footnotesize 
	\begin{tabular}{ccc}
		\hline
		& 
		training set& 
		test set \\
		\hline
		The aligned cytopathology dataset& 2,257 & 966 \\
		The aligned histopathology dataset& 19,200 & 7,936\\
		\hline
	\end{tabular}
\end{table} 

\begin{table*}[!h]
	\footnotesize 
	\centering
	\caption{The number of image patches on the cytopathology and histopathology classification dataset. The classification set is used to train and test the classifier and the stain normalization set is used to train the stain normalization methods.}
	\label{table2}
	%	\setlength{\tabcolsep}{4pt}
	%\footnotesize 
	\begin{tabular}{ccccccc}
		\hline
		The cytopathology classification dataset &D1& D2& D3& D4& D5 &\\
		\hline
		The classification set& 244,341& 10,000& 10,000& 10,000& 10,000 & \\
		The stain normalization set& 2,500& 625& 625& 625& 625 &\\
		\hline
		The histopathology classification dataset & Uni16& C1& C2& C3& C4& C5\\
		\hline
		The classification set& 40,000&13,613&12,355&14,076&9,235&9,696\\
		The stain normalization set& 6,000&1,200&1,200&1,200&1,200&1,200 \\
		\hline
	\end{tabular}
\end{table*}

\subsubsection{Cytopathology Classification Dataset}
The cytopathology classification dataset was used to verify the performance of the stain normalization algorithm on a multi-domain cytopathology classification diagnostic task. The slides in the cytopathology classification dataset are from Hubei Provincial Maternal and Child Health Hospital and Tongji Hospital. There are five different styles (D1-D5), among which D1-D4 come from Hubei Maternal and Child Health Hospital, and D5 comes from Tongji Hospital. In this dataset, the image patches that contain abnormal cells are labeled as abnormal patches, and the image patches that do not contain any abnormal cells are labeled as normal patches. There are 244,341 image patches in D1 and 40,000 image patches in D2-D5 (10,000 image patches in each center) as shown in Table \ref{table2}. We randomly picked 2,500 image patches in D1 as the target images and 2,500 image patches in D2-D5 as the source images for training the stain normalization algorithm. For the classifier, we used D1 to train the classifier, and used D2-D5 to test the performance of the classifier with and without normalization.

\subsubsection{Histopathology Classification Dataset}
The histopathology classification datasets used the publicly available Camelyon16 \citep{camelyon16} (399 whole slide images from two centers) and Camelyon17 \citep{camelyon17} datasets (1,000 whole slide images from five centers). In our experiments, 100 WSIs from University Medical Center Utrecht (Uni16) in Camelyon16 training part were used to extract the training patches, and 500 WSIs from the five centers (C1-C5) in Camelyon17 training part were used to extract the test patches. In this dataset, the image patches that contain tumor cells are labeled as abnormal patches, and the image patches that do not contain any tumor cells are labeled as normal patches. There are 40,000 image patches from Uni16 in the training set of this dataset. And there are 58,975 image patches in the test set of this dataset as shown in Table \ref{table2}. We randomly picked 6,000 image patches from the training set as the target images and 6,000 image patches from the test set as the source images for training stain normalization algorithms. For the classifier, we used the training set to train the classifier, and used the test set to test the performance of the classifier with and without normalization.

\subsection{Evaluation Metrics}
In order to evaluate different methods fairly and effectively, we considered in four aspects: similarity with target image, preservation of source image information, computational efficiency, and classifier accuracy.

In order to effectively measure the similarity to the target image and preserve the information of the source image, we used three similarity matrices, Quaternion Structural Similarity (QSSIM) \citep{qssim}, Structural Similarity index (SSIM) \citep{ssim} and Peak Signal-to-Noise Ratio (PSNR). The similarity between the normalized image and the target image was evaluated using QSSIM, SSIM and PSNR, denoted as QSSIM Target, SSIM Target and PSNR Target. The preservation of source image information was evaluated using SSIM between the normalized  image and the source image, denoted as SSIM Source. SSIM Target and PSNR Target were calculated using raw RGB values. Similar to [11,27], SSIM Source was calculated using grayscale images. The statistic results of SSIM Target, PSNR Target, and SSIM Source on the testing set in the aligned cytopathology dataset and the aligned histopathology dataset are shown in Tables \ref{table3} and \ref{table4}, as mean ± standard deviation. 

In order to evaluate the computational efficiency of different methods, the number of frames per second (FPS) was calculated on the system with 6-core Intel(R) Core (TM) i7-6850K CPU and NVidia GeForce GTX 1080Ti. The input and output (IO) time was not included. The FPS results of different methods are shown in Table \ref{table5} when the input image size is 512×512 pixels.

The accuracy was used to evaluate the classifier performance. The statistic results of accuracy on the cytopathology and histopathology classification datasets are shown in Tables \ref{table6} and \ref{table7}, as mean ± standard deviation.

\begin{figure*}[!h]
	\footnotesize 
	\centering
	\includegraphics[width=154 mm]{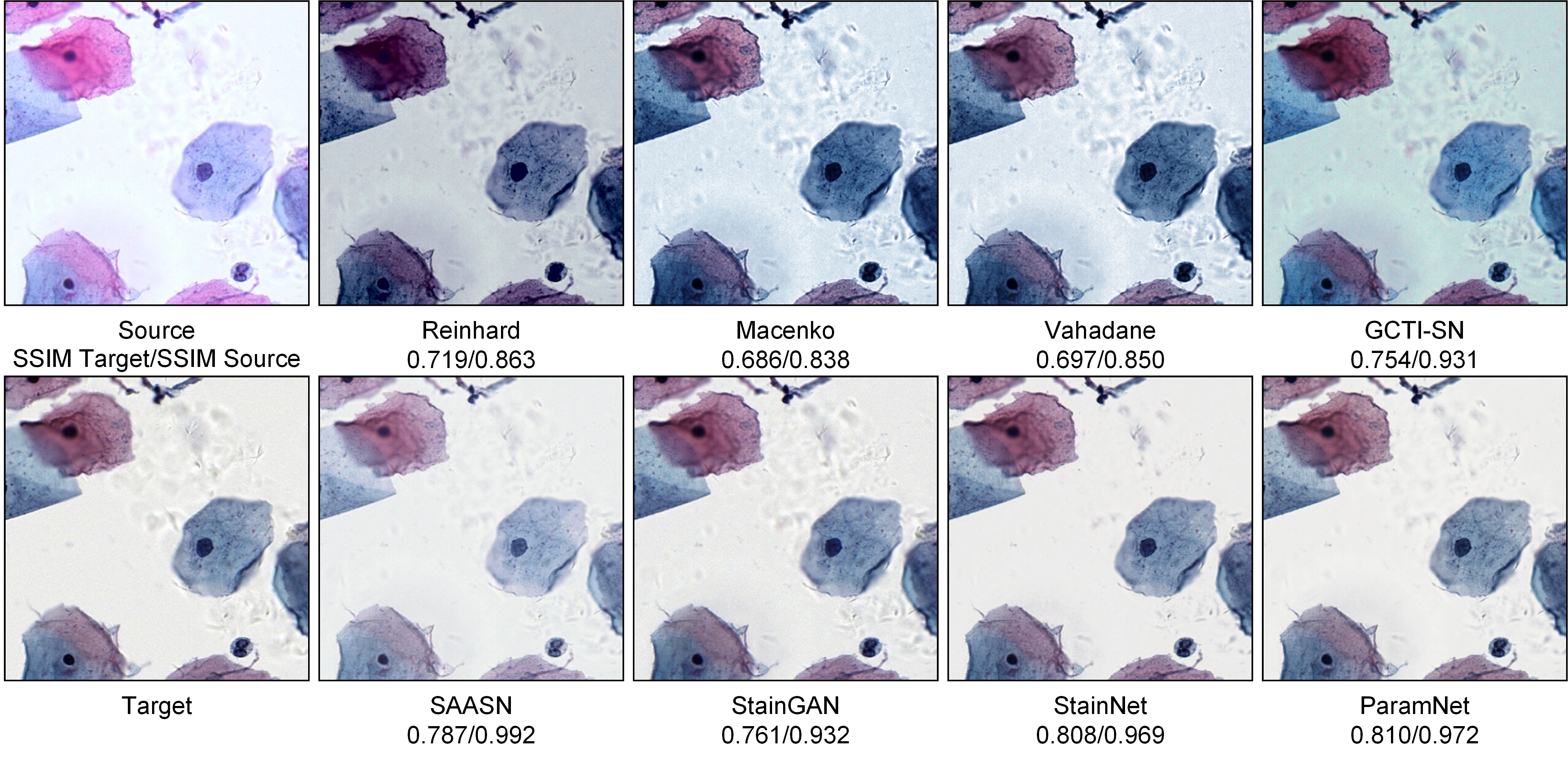}
	\caption{Visual comparison among different methods on the aligned cytopathology dataset.}
	\label{fig3}
\end{figure*}

\subsection{Implementation}

For conventional methods (Reinhard \citep{reinhard2001color}, Macenko \citep{macenko2009method}, Vahadane \citep{vahadane2016structure} and GCTI-SN \citep{gupta2020gcti}), one professionally selected image was used as the reference image. For the GAN based methods (SAASN \citep{shrivastava2021self}, StainGAN \citep{shaban2019staingan}, and StainNet \citep{kang2021stainnet}), we used their default training settings.

For ParamNet, four losses are included in the training framework. During training, we randomly transformed image resolutions (0.5$\sim$1.0) to stabilize learning. We used Adam \citep{kingma2014adam} to optimize the network, and the learning rate is linearly increased from 0 to 2e-4 in the first 1,000 iterations, and then linearly decreased to 0 until the end of training.  We trained ParamNet for 200k iterations on the aligned cytopathology dataset and aligned histopathology dataset, and 400k iterations on the cytopathology classification dataset and histopathology classification dataset.

For the classifier, we used SqueezeNet \citep{squeezenet} pre-trained on ImageNet \citep{imagenet} as the classification network.  The classifier was trained using cross entropy loss and Adam \citep{kingma2014adam} optimizer. The initial learning rate was set to 0.0002, and it reduced by a factor of 0.1 at the 40th and 50th epoch. The training was stopped at the 60th epoch, which was chosen experimentally. The experiment was repeated 10 times in order to enhance reliability.
\subsection{Results}
In this section, we compared our method with state-of-art normalization methods, including Reinhard \citep{reinhard2001color}, Macenko \citep{macenko2009method}, Vahadane \citep{vahadane2016structure}, GCTI-SN \citep{gupta2020gcti}, SAASN \citep{shrivastava2021self}, StainGAN \citep{shaban2019staingan}, and StainNet \citep{kang2021stainnet}.
\subsubsection{One-to-One Stain Normalization}

\begin{figure*}[!h]
	\footnotesize 
	\centering
	\includegraphics[width=154 mm]{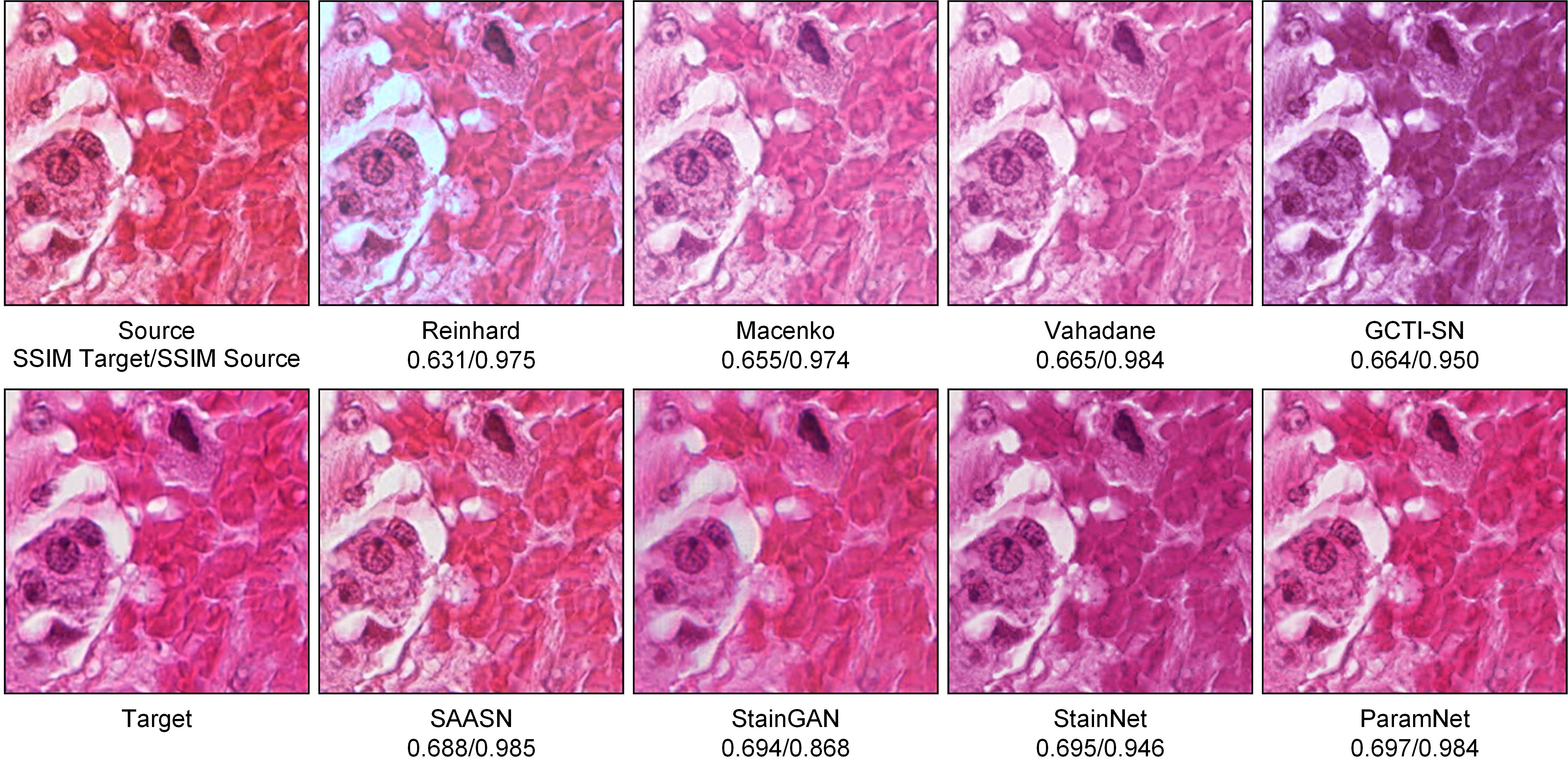}
	\caption{Visual comparison among different methods on the aligned histopathology dataset.}
	\label{fig4}
\end{figure*}

\begin{table*}[!h]
	\footnotesize 
	\centering
	\caption{Different evaluation metrics are reported for various stain normalization methods on the aligned cytopathology dataset.}
	\label{table3}
	%	\footnotesize 
	\begin{tabular}{ccccc}
		\hline
		Methods  & QSSIM Target & SSIM Target & PSNR Target & SSIM Source \\ \hline
		Source  & 0.825±0.027  & 0.765±0.028 & 16.9±1.89  & 1.000±0.000 \\
		Reinhard & 0.814±0.051  & 0.739±0.046 &  19.8±3.3   & 0.885±0.042 \\
		Macenko  & 0.778±0.064  & 0.731±0.054 &  22.5±3.1   & 0.853±0.054 \\
		Vahadane & 0.784±0.059  & 0.739±0.050 &  22.6±3.0   & 0.867±0.050 \\
		GCTI-SN  & 0.832±0.039  & 0.764±0.039 &  19.7±2.7   & 0.905±0.034 \\
		SAASN   & 0.855±0.025  & 0.778±0.026 &  21.5±2.0   & \textbf{0.990±0.001} \\
		StainGAN & 0.832±0.029  & 0.764±0.030 &  29.7±1.6   & 0.905±0.021 \\
		StainNet & 0.874±0.027  & 0.809±0.027 &  29.8±1.7   & 0.945±0.025 \\
		ParamNet & \textbf{0.878±0.026}  & \textbf{0.816±0.026} &  \textbf{30.0±1.6}   & 0.951±0.023 \\ \hline
	\end{tabular}
\end{table*} 

\begin{table*}[!h]
	\footnotesize 
	\centering
	\caption{Different evaluation metrics are reported for various stain normalization methods on the aligned histopathology dataset.}
	\label{table4}
	%	\footnotesize 
	\begin{tabular}{ccccc}
		\hline
		Methods  & QSSIM Target & SSIM Target & PSNR Target & SSIM Source \\ \hline
		Source  & 0.690±0.099  & 0.641±0.108 &  20.3±3.2   & 1.000±0.000 \\
		Reinhard & 0.682±0.092  & 0.617±0.106 &  19.9±2.1   & 0.964±0.031 \\
		Macenko  & 0.703±0.101  & 0.656±0.115 &  20.7±2.7   & 0.966±0.049 \\
		Vahadane & 0.708±0.102  & 0.664±0.116 &  21.1±2.8   & 0.967±0.046 \\
		GCTI-SN  & 0.685±0.093  & 0.638±0.100 &  19.7±2.2   & 0.947±0.043 \\
		SAASN   & \textbf{0.739±0.098}  & 0.690±0.114 &  21.6±3.5   & \textbf{0.994±0.002} \\
		StainGAN & 0.733±0.088  & \textbf{0.706±0.099} &  \textbf{22.7±2.6}   & 0.912±0.025 \\
		StainNet & 0.724±0.097  & 0.691±0.107 &  22.5±3.3   & 0.957±0.007 \\
		ParamNet & 0.720±0.098  & 0.687±0.106 &  \textbf{22.7±3.0 }  & 0.966±0.001 \\ \hline
	\end{tabular}
\end{table*}

\begin{table*}[!h]
	\footnotesize 
	\centering
	\caption{Quantitative comparison of computational efficiency among different methods. The input and output (IO) time is not included. The input image size is 512×512 pixels.}
	\label{table5}
	%	\setlength{\tabcolsep}{4pt}
	%\footnotesize 
	\begin{tabular}{ccccccccc}
		\hline
		Methods& Reinhard& Macenko& Vahadane& GCTI-SN& SAASN& StainGAN& StainNet& \textbf{ParamNet} \\
		\hline
		FPS& 54.8& 4.0& 0.5& 6.4& 5.2& 19.6& 881.8& \textbf{1605.2} \\
		\hline
	\end{tabular}
\end{table*}

We compared our ParamNet with state-of-art normalization methods on the one-to-one stain normalization task. We reported the following results on the aligned cytopathology dataset and the aligned histopathology dataset: (1) Visual comparison among different methods. (2) Quantitative comparison of similarities among different methods. (3) Quantitative comparison of computational efficiency among different methods.

\textbf{Visual comparison}

Firstly, we compared our ParamNet with other stain normalization methods in the visual appearance. The visual comparison among different methods is shown in Fig. \ref{fig3} and \ref{fig4}.

The results on the aligned cytopathology dataset are shown in Fig. \ref{fig3}. For conventional methods (Reinhard, Macenko, Vahadane and GCTI-SN), the normalized images are far from the target image in color and brightness. The conventional methods only extract information from a single reference image. Moreover, it is difficult to extract the precise color mapping relationship in this way, resulting in poor performance of the normalized results. For the image normalized by SAASN, its brightness is closer to the source image than to the target image. The images normalized by StainGAN, StainNet and ParamNet are very close to the target image. However, in terms of image content, the image normalized by StainGAN has a more complex background, while the image StainNet has a simpler background, only the image normalized by ParamNet is closest to the source image. StainGAN uses a complex convolutional network that tends to introduce additional noise. StainNet uses a single 1×1 convolutional network, which makes it not capable enough to handle complex color mapping relationships. ParamNet, on the other hand, can automatically determine the best parameters (weights and biases of convolutional layers) for each input image, thus achieving better performance.

The results on the aligned histopathology dataset are shown in Fig. \ref{fig4}. For conventional methods (Reinhard, Macenko, Vahadane and GCTI-SN), there are still some differences between the normalized image and the target image in color and brightness. However, the performance of conventional methods on aligned histopathology datasets is significantly better than that on aligned cytopathology datasets. This is because most of the conventional methods are designed for histopathology images and are not suitable for cytopathology images. For the image normalized by SAASN, its color is lighter to the target image. StainGAN introduces blur to the normalized image. The image normalized by StainNet is darker than the target image. Only the image normalized by ParamNet is most close to the target image.

\textbf{Quantitative comparison of similarities}

Secondly, we quantitatively compared our ParamNet with other stain normalization methods in the similarity to the target image and the similarity to the source image. The quantitative comparison among different methods is shown in Table \ref{table3} and \ref{table4}.

The results on the aligned cytopathology dataset are shown in Table \ref{table3}. The conventional methods (Reinhard, Macenko, Vahadane and GCTI-SN) have low QSSIM Target, SSIM Target and PSNR Target, which means that there is a large difference between the images normalized by the conventional methods and the target images. SAASN has the highest SSIM Source but low PSNR Target, which means that it preserves the information of the source images as much as possible but ignores the similarity with the target images. But the purpose of stain normalization is to be similar to the target images, so SAASN is inappropriate for stain normalization. StainGAN, StainNet and ParamNet have high PSNR Target, which means that the images normalized by them are similar enough to the target images. Among these three methods, StainGAN has the lowest SSIM Source, which means it may lose some information of the source images. For our ParamNet, it has the highest PSNR Target and also has a high SSIM Source, which means that it can achieve a high similarity to the target images while preserving the source images information as much as possible.

The results on the aligned histopathology dataset are shown in Table \ref{table4}. The conventional methods (Reinhard, Macenko, Vahadane and GCTI-SN) have lower QSSIM Target, SSIM Target and PSNR Target than the GAN based methods (SAASN, StainGAN, StainNet and ParamNet) in Table \ref{table4}, which means that the GAN based methods can achieve better performance. Among the GAN based methods, SAASN has the highest SSIM Source and the lowest PSNR Target, StainGAN and ParamNet have the highest PSNR Target, and ParamNet has the second highest SSIM Source. So, our ParamNet can achieve a high similarity to the target images while preserving the source images information as much as possible on the aligned histopathology dataset.

\textbf{Quantitative comparison of computational efficiency}

Finally, we quantitatively compared the computational efficiency of our method with other stain normalization methods. We tested the FPS of different methods with an input image size of 512×512 pixels and show the results in Table \ref{table5}. As shown in Table \ref{table5}, our ParamNet (1,605.2 FPS) is much faster than StainNet (881.8 FPS), which means that our ParamNet can normalize an 100,000×100,000 whole slide image (WSI) within 25s. The prediction sub-network of ParamNet computes at low resolution and the color mapping sub-network of ParamNet computes at high resolution. The sub-network has only two 1×1 convolutional layers with 8 channels, however, StainNet has three 1×1 convolutional layers with 32 channels. So our ParamNet is computationally more efficient than StainNet.

%\begin{table}
%	\centering
%	\caption{Quantitative comparison of computational efficiency among different methods. The input and output (IO) time is not included. The input image size is 512×512.}
%	\label{table5}
%	\setlength{\tabcolsep}{3pt}
%	\begin{tabular}{|p{40pt}|p{40pt}|p{40pt}|p{40pt}|p{40pt}|}
	%			\hline
	%			Methods& Reinhard& Macenko& Vahadane& GCTI-SN\\
	%			\hline
	%			FPS& 54.8& 4.0& 0.5& 6.4 \\
	%			\hline
	%			Methods& SAASN& StainGAN& StainNet& ParamNet \\
	%			\hline
	%			FPS & 5.2& 19.6& 881.8& 1605.2\\ 
	%			\hline
	%		\end{tabular}
%\end{table}
\subsubsection{Multi-to-One Stain Normalization}
\begin{figure*}[!h]
	\footnotesize 
	\centering
	\includegraphics[width=170 mm]{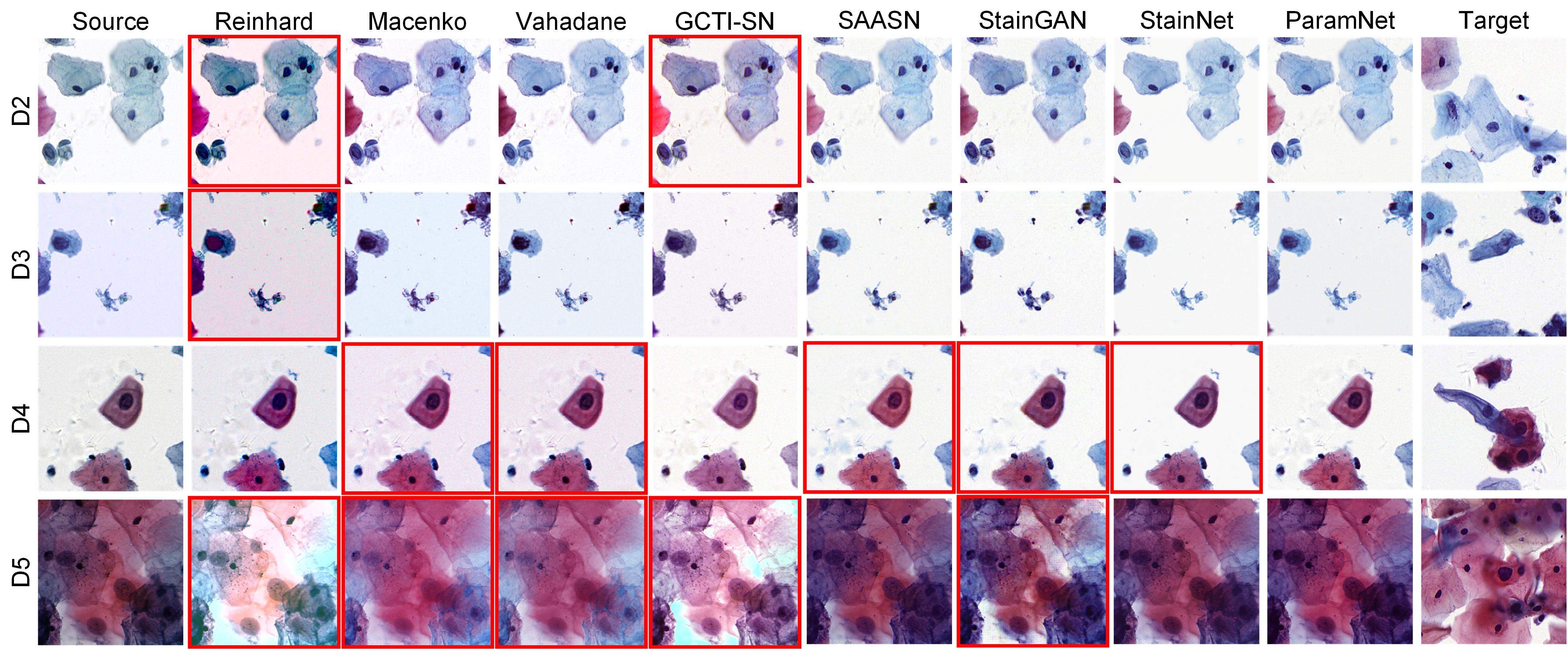}
	\caption{Visual comparison among different methods on the cytopathology classification dataset. The source images have multiple styles (D2-D5), and the target images are from D1. Red boxes mark obviously abnormal normalized results.}
	\label{fig5}
\end{figure*}

\begin{table*}[!h]
	\footnotesize 
	\centering
	\caption{SSIM Source for various stain normalization methods on the cytopathology classification dataset.}
	\label{table8}
	%	\footnotesize 
	\begin{tabular}{cccccc}
		\hline
		Methods  &     D2      &     D3      &     D4      &     D5      &   Average   \\ \hline
		Reinhard & 0.914±0.125 & 0.872±0.198 & 0.811±0.171 & 0.815±0.193 & 0.853±0.172 \\
		Macenko  & 0.866±0.169 & 0.857±0.230 & 0.766±0.203 & 0.727±0.192 & 0.804±0.198 \\
		Vahadane & 0.848±0.189 & 0.850±0.247 & 0.751±0.205 & 0.715±0.199 & 0.791±0.210 \\
		GCTI-SN  & 0.905±0.175 & 0.901±0.199 & 0.893±0.150 & 0.878±0.181 & 0.894±0.176 \\
		SAASN   & 0.992±0.003 & 0.992±0.002 & 0.986±0.006 & 0.983±0.006 & 0.988±0.004 \\
		StainGAN & 0.954±0.035 & 0.904±0.077 & 0.896±0.062 & 0.874±0.051 & 0.907±0.056 \\
		StainNet & 0.941±0.032 & 0.968±0.012 & 0.881±0.040 & 0.898±0.029 & 0.922±0.028 \\
		\textbf{ParamNet} & \textbf{0.993±0.010} & \textbf{0.994±0.001} & \textbf{0.989±0.006} & \textbf{0.987±0.008} & \textbf{0.991±0.006} \\ \hline
	\end{tabular}
\end{table*}

\begin{figure*}
	\footnotesize 
	\centering
	\includegraphics[width=170 mm]{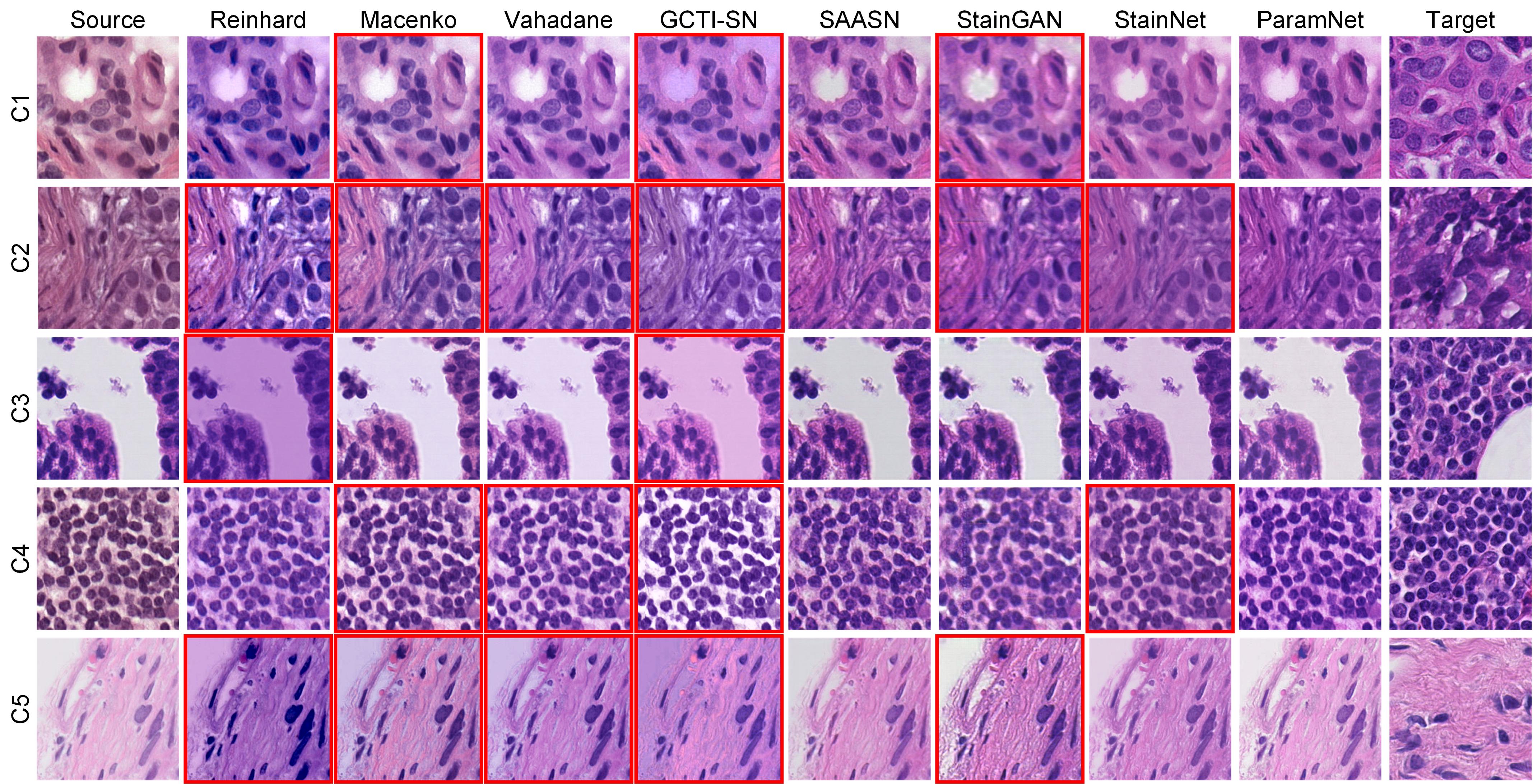}
	\caption{Visual comparison among different methods on the histopathology classification dataset. The source images are from five centers (C1-C5) of Camelyon17, and the target images are from Uni16 of Camelyon16. The source images of C3 and the target images are from same center but at different times. Red boxes mark obviously abnormal normalized results.}
	\label{fig6}
\end{figure*}

Firstly, we compared our ParamNet with other stain normalization methods in the visual appearance as shown in Fig. \ref{fig5} and \ref{fig6}.

The results on the cytopathology classification dataset are shown in Fig. \ref{fig5}. As can be seen, there are many wrong variations in color and brightness on the normalized images by the conventional methods (Reinhard, Macenko, Vahadane and GCTI-SN). For example, the cell in the source image of D2 is incorrectly normalized from dark red to bright red by GCTI-SN, and the brightness of the D5’s source image is over-enhanced by Reinhard and GCTI-SN. StainGAN and SAASN use complex convolutional networks for stain normalization, which not only reduces the computational efficiency of the algorithm but also brings artifacts to the normalized image. For example, SAASN and StainGAN add extra contents to the normalized images that do not belong to the source images in D4. And StainGAN removes some contents belonging to the source image in the normalized image of D5. As for StainNet, it has a single three-layer 1×1 convolutional network, which limits its color-mapping capability. So, there are many details missing, especially on the background of the normalized image by StainNet. Unlike StainNet, our ParamNet uses a dynamic parameter 1×1 convolutional network for color mapping, so our ParamNet performs well on the multi-to-one stain normalization task.

The results on the histopathology classification dataset are shown in Fig. \ref{fig6}. As can be seen, the conventional methods perform better on histopathology images, but there are still many wrong variations in color and brightness. For example, the light background in the source image of center3 is incorrectly normalized to the purple background by Reinhard and GCTI-SN, and the brightness of the center5’s source image is over-enhanced by Macenko and Vahadane. The GAN based methods all show good performance. However, the normalized images by StainGAN are a little blurry and the normalized images by StainNet are a bit purple. Our ParamNet can achieve the best performance among the GAN based methods.

Secondly, we compared our ParamNet with other stain normalization methods using SSIM Source as shown in Table \ref{table8} and \ref{table9}.

Whether on the histopathology classification dataset or the cytopathology classification dataset, our ParamNet obtained the highest SSIM Source, which shows that our ParamNet retains the source image information best on the two datasets. StainNet uses a fixed parameter 1×1 convolutional network to perform normalization, whereas our ParamNet uses a dynamic parameter 1×1 convolutional network. The feature of dynamic parameters enable our network to use the best network parameters for each input image, so as to better retain the information of the source image. This is why the SSIM Source of our method can be better than StainNet.

\begin{table*}[!h]
	\footnotesize 
	\centering
	\caption{SSIM Source for various stain normalization methods on the histopathology classification dataset.}
	\label{table9}
	%	\footnotesize 
	%	\setlength{\tabcolsep}{3pt}
	\begin{tabular}{ccccccc}
		\hline
		Methods  &     C1      &     C2      &     C3      &     C4      &     C5      &   Average   \\ \hline
		Reinhard & 0.852±0.079 & 0.907±0.079 & 0.881±0.088 & 0.863±0.101 & 0.726±0.099 & 0.846±0.089 \\
		Macenko  & 0.902±0.084 & 0.722±0.152 & 0.939±0.064 & 0.918±0.085 & 0.803±0.092 & 0.857±0.096 \\
		Vahadane & 0.930±0.078 & 0.843±0.159 & 0.941±0.065 & 0.931±0.081 & 0.819±0.087 & 0.893±0.094 \\
		GCTI-SN  & 0.835±0.148 & 0.887±0.103 & 0.899±0.076 & 0.867±0.127 & 0.726±0.095 & 0.843±0.110 \\
		SAASN   & 0.974±0.039 & 0.981±0.013 & \textbf{0.989±0.005} & 0.975±0.018 & 0.953±0.017 & 0.974±0.018 \\
		StainGAN & 0.933±0.047 & 0.952±0.019 & 0.966±0.019 & 0.928±0.037 & 0.869±0.053 & 0.930±0.035 \\
		StainNet & 0.945±0.039 & 0.947±0.016 & 0.886±0.031 & 0.947±0.018 & 0.898±0.017 & 0.925±0.024 \\
		ParamNet & \textbf{0.974±0.038} & \textbf{0.988±0.011} & 0.983±0.019 & \textbf{0.976±0.023} & \textbf{0.958±0.066} & \textbf{0.976±0.032} \\ \hline
	\end{tabular}
\end{table*}

\subsubsection{Application results on the classification task}

\begin{table*}[!h]
	\footnotesize 
	\centering
	\caption{Accuracy for various stain normalization methods on the cytopathology classification dataset.}
	\label{table6}
	%	\footnotesize 
	%	\setlength{\tabcolsep}{3pt}
	\begin{tabular}{cccccc}
		\hline
		Accuracy& D2& D3& D4& D5& Average\\
		\hline
		Original& 0.853±0.031& 0.915±0.022& 0.873±0.029& 0.688±0.033& 0.832±0.029\\
		Reinhard& 0.862±0.015& 0.867±0.025& 0.796±0.026& 0.728±0.025& 0.813±0.023\\
		Macenko& 0.929±0.014& 0.937±0.023& 0.915±0.010& 0.818±0.034& 0.900±0.020\\
		Vahadane& 0.905±0.022& 0.940±0.031& 0.907±0.019& 0.745±0.037& 0.875±0.026\\
		GCTI-SN& 0.932±0.016& 0.944±0.022& 0.924±0.017& 0.850±0.039& 0.912±0.024\\
		SAASN& 0.964±0.013& 0.911±0.024& \textbf{0.944±0.010}& 0.863±0.039& 0.920±0.022\\
		StainGAN& 0.918±0.007& 0.908±0.011& 0.867±0.006& 0.799±0.018& 0.873±0.011\\
		StainNet& 0.928±0.034& 0.963±0.014& 0.924±0.032& 0.820±0.059& 0.909±0.035\\
		ParamNet& \textbf{0.968±0.013}& \textbf{0.965±0.017}& 0.943±0.006& \textbf{0.867±0.039}& \textbf{0.936±0.019}\\
		\hline
	\end{tabular}
\end{table*}

\begin{table*}[!h]
	\footnotesize 
	\centering
	\caption{Accuracy for various stain normalization methods on the histopathology classification dataset.}
	\label{table7}
	%	\footnotesize 
	%	\setlength{\tabcolsep}{3pt}
	\begin{tabular}{ccccccc}
		\hline
		Accuracy& C1& C2& C3& C4& C5& Average\\
		\hline
		Original& 0.500±0.016& 0.561±0.030& 0.907±0.003& 0.528±0.014& 0.765±0.041& 0.652±0.021\\
		Reinhard& 0.871±0.014& 0.855±0.008& 0.815±0.006& 0.823±0.027& 0.808±0.015& 0.834±0.014\\
		Macenko& 0.814±0.009& 0.832±0.009& 0.816±0.007& 0.836±0.012& 0.789±0.010& 0.817±0.009\\
		Vahadane& 0.817±0.005& 0.826±0.005& 0.762±0.007& 0.890±0.004& 0.805±0.003& 0.820±0.005\\
		GCTI-SN& 0.840±0.006& 0.832±0.007& 0.816±0.007& 0.879±0.005& 0.804±0.006& 0.834±0.006\\
		SAASN& 0.883±0.004& \textbf{0.887±0.002}& 0.908±0.003& 0.888±0.004& 0.819±0.010& 0.877±0.005\\
		StainGAN& 0.898±0.004& 0.871±0.001& 0.921±0.003& 0.903±0.004& 0.881±0.005& 0.895±0.003\\
		StainNet& 0.898±0.006& 0.836±0.006& 0.895±0.005& 0.862±0.019& 0.877±0.005& 0.874±0.008\\
		ParamNet& \textbf{0.920±0.003}& 0.857±0.007& \textbf{0.924±0.003}& \textbf{0.899±0.004}& \textbf{0.884±0.003}& \textbf{0.897±0.004}\\
		\hline
	\end{tabular}
\end{table*}
Moreover, we compared our ParamNet with other stain normalization methods on the classification task. The accuracy for various stain normalization methods is shown in Table \ref{table6},\ref{table7}. 

The accuracy for various stain normalization methods on the cytopathology classification dataset is shown in Table \ref{table6}. The classifier has poor performance on the original image of D2-D5, and the classifier has different performance improvements on the images normalized by different methods (except Reinhard). Among conventional methods, the classifier has the best accuracy (0.912 on average) on the images normalized by GCTI-SN. And among GAN based methods, the classifier has the best accuracy (0.936 on average) on the images normalized by our ParamNet. StainGAN and StainNet do not get competitive performance on the cytopathology classification dataset.

\begin{figure*}[!h]
	\footnotesize 
	\centering
	\includegraphics[width=170 mm]{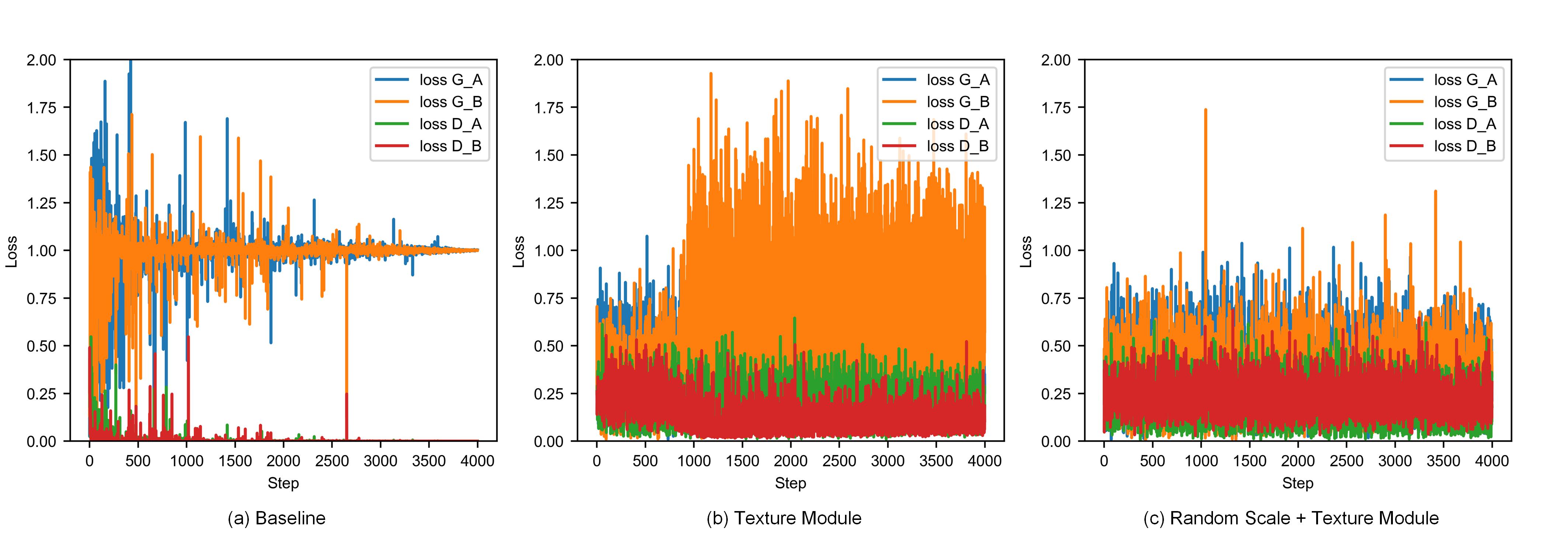}
	\caption{GAN loss during training with different strategies. (a) The baseline framework’s stability of learning is poor. (b) The stability of learning is enhanced after the texture module is introduced in the framework. (c) The combination of texture module and random scale can effectively enhance the stability of learning.}
	\label{fig7}
\end{figure*}

\begin{table*}[!h]
	\footnotesize 
	\centering
	\caption{Accuracy of models with different tricks on the histopathology classification dataset.}
	\label{table10}
	%	\setlength{\tabcolsep}{1pt}
	%	\footnotesize 
	\begin{tabular}{ccccccc}
		\hline
		Model & C1 & C2 & C3 & C4 & C5 & Average \\
		\hline
		Baseline & 0.848±0.012 & \textbf{0.884±0.005} & 0.780±0.025 & 0.785±0.027 & 0.817±0.028 & 0.823±0.019 \\
		+Texture Module & \textbf{0.921±0.004} & 0.860±0.006 & 0.811±0.019 & 0.900±0.004 & 0.848±0.010 & 0.868±0.009 \\
		+Random Scale & 0.917±0.003 & 0.855±0.006 & 0.859±0.009 & 0.899±0.006 & 0.846±0.014 & 0.875±0.008 \\
		+Identity Loss & 0.916±0.003 & 0.857±0.006 & 0.899±0.004 & 0.901±0.003 & \textbf{0.885±0.003} & 0.892±0.004 \\
		+Domain Consistency Loss & 0.920±0.003 & 0.857±0.006 &\textbf{ 0.924±0.003} & \textbf{0.899±0.004} & 0.884±0.003 & \textbf{0.897±0.004} \\
		\hline
	\end{tabular}
\end{table*}

The accuracy for various stain normalization methods on the histopathology classification dataset is shown in Table \ref{table7}. The classifier has poor performance on the original images of C1-C5, even the accuracy of the classifier is close to 0.5 in C1, C2 and C4. Overall, using the normalized images can improve the performance of the classifier compared to the original images. On the histopathology classification dataset, the classifier has better performance on images normalized by the GAN-based methods than the conventional methods. Among all the normalization methods in Table \ref{table7}, our method has the highest accuracy (0.897 on average). The images in C3 and Uni16 are from the same medical center but at different times. For the classifier trained on Uni16, the accuracy is 0.907 on the original images of C3. And the accuracy can be increased to 0.924 on the normalized images by our ParamNet, which shows that our method can improve the accuracy of the classifier, even on images from the same center.

\subsubsection{Ablation study}
To verify the effectiveness of our proposed training framework, an ablation study was performed on the histopathology classification dataset. The accuracy of models trained with different strategies is shown in Table \ref{table10}. In addition, we also visualize the changes in GAN loss during the training process as shown in Fig. \ref{fig7}. During the training, GAN loss is recorded every 100 iterations.

As can be seen from Fig. \ref{fig7} (a), for the baseline model, the discriminator loss is almost 0, which shows that the discriminator can almost completely distinguish between real images and fake images. But for the least squares loss in our training framework, it is difficult for the discriminator to provide effective gradients to optimize the generator in this case. In order to solve this problem, a texture module is first introduced between the generator and the discriminator. As can be seen from Fig. \ref{fig7} (b), although it helps to solve the problem of imbalance between the generator and the discriminator, there is still a slight imbalance. After introducing the texture module, we introduce random scales during the training process to make the discriminator focus more on the overall color difference. As can be seen from Fig. \ref{fig7} (c), the combined use of random scale and texture modules can well solve the problem of imbalance between the generator and the discriminator.

The accuracy of models trained using different strategies on the histopathology classification dataset is given in Table \ref{table10}, where the average accuracy of the baseline model is only 0.823. The texture modules and random scale are used to solve the problem of generator and discriminator imbalance. Combining the texture module with random scales increases the average accuracy to 0.875. Identity loss is used to enhance domain-invariant features, and the accuracy can be improved by 0.017. Domain consistency loss can enhance the consistency of the output of the generator and texture module, and can improve the accuracy from 0.892 to 0.897. Through the combination of the four strategies proposed in this article, the accuracy is finally increased from 0.823 to 0.897.

\section{Discussion and Conclusion}
In this paper, we proposed a novel stain normalization network, ParamNet, and an adversarial training framework. ParamNet contains two sub-networks, a modified ResNet18 \citep{he2016deep} as the prediction sub-network and a fully 1×1 convolutional network as the color mapping sub-network, where the prediction sub-network can predict all parameters (weights and biases of convolutional layers) of the color mapping sub-network at a low resolution and the color mapping sub-network can directly normalize the source images at the original resolution. Moreover, we validated the effectiveness of our method on four datasets, the results show that ParamNet has better performance than other methods.

The feature of dynamic parameter ensures that our network has a sufficient capability for multi-domain normalization. And the color mapping sub-network has an extremely concise network structure, which allows our network to be computationally efficient and structurally consistent. 
%However, since different input images correspond to different parameters, there will be slight inconsistencies between adjacent image blocks. We hope to solve this problem in further study.

\section*{Acknowledgments}
This work is supported by the NSFC project (grant 62375100, 62201221), China Postdoctoral Science Foundation (grant 2021M701320, 2022T150237) and Southern Medical University Research Start-up Foundation. 
\bibliographystyle{elsarticle-num-names.bst}
\biboptions{authoryear}
\bibliography{reference.bib}
\end{document}